\documentclass[aip,jap,reprint,amsmath,amssymb]{revtex4-2}
\usepackage[T1]{fontenc}
\usepackage{graphicx}
\usepackage{bm}
\usepackage{url}
\usepackage{xcolor}
\begin{document}
\title{DefectSkills: Guiding AI agents through first-principles calculations for point defects in semiconductors and insulators}
\author{Zhenkun Yuan}
\email{zkyuans@gmail.com}
\author{Yizhi Zhu}
\author{Geoffroy Hautier}
\email{gh55@rice.edu}
\affiliation{Department of Materials Science and NanoEngineering, Rice University, Houston, TX 77005, USA}
\affiliation{Rice Advanced Materials Institute, Rice University, Houston, TX 77005, USA}
\date{\today}
\begin{abstract}
First-principles defect calculations can now be performed autonomously by AI agents, enabled by established computational methods and open-source Python toolkits. However, reliable and repeatable defect calculations require clear scientific and technical guidance that helps agents resolve ambiguities and avoid errors in calculation setup, execution, postprocessing, and analysis. Here, we present DefectSkills, a set of structured prompts that provides human-in-the-loop guidance to AI agents performing first-principles calculations of point defects in semiconductors and insulators using the Vienna Ab initio Simulation Package (VASP). The prompts encode human expertise and guide agents through unit-cell relaxation, electronic-structure and dielectric calculations, phase-stability and chemical-potential analysis, charged-defect supercell generation and calculations, finite-size corrections, formation-energy calculations and thermodynamic analysis, and data backup. By packaging this expertise as reusable agent skills, DefectSkills supports consistent workflow execution across agentic AI systems and reproducible defect calculations.
\end{abstract}
\maketitle
\raggedbottom
\setlength{\parskip}{0pt}
\let\originalitemize\itemize
\let\endoriginalitemize\enditemize
\renewenvironment{itemize}
  {\originalitemize
   \setlength{\itemsep}{0pt}
   \setlength{\parsep}{0pt}
   \setlength{\parskip}{0pt}}
  {\endoriginalitemize}
\setlength{\abovedisplayskip}{6pt plus 2pt minus 2pt}
\setlength{\belowdisplayskip}{6pt plus 2pt minus 2pt}
\setlength{\abovedisplayshortskip}{3pt plus 1pt}
\setlength{\belowdisplayshortskip}{3pt plus 1pt minus 1pt}
\setlength{\bibsep}{0pt}

\section{Introduction}

First-principles calculations are essential for investigating point defects in materials.\cite{Freysoldt2014} They provide microscopic insight into the atomic and electronic structures of defects and clarify how defects influence material properties. These insights, particularly those obtained using hybrid density-functional theory (DFT),\cite{Krukau2006} can guide defect control and doping strategies and are increasingly used to design defects and materials with targeted properties.\cite{Weber2010,YuanJoule2024}

However, first-principles defect studies involve multiple interconnected stages, including unit-cell relaxation, electronic-structure and dielectric calculations, phase-stability and chemical-potential analysis, defect generation and supercell calculations, finite-size corrections, formation-energy calculations, and thermodynamic analysis. Choices made at one stage can propagate through subsequent stages, and variations in calculation setup, execution, and analysis can lead to substantial discrepancies in predicted defect properties.\cite{Squires2026Guidelines} For example, studies of BaZrS$_3$ have reported differing defect energetics and assessments of dopability and defect tolerance, despite agreement on some qualitative trends.\cite{Meng2016,Wu2021,Yuan2024,Desai2025} Established open-source Python toolkits, including PyCDT,\cite{Broberg2018} pydefect,\cite{Kumagai2021} doped,\cite{Kavanagh2024} and \texttt{pymatgen-analysis-defects},\cite{Shen2024} have improved the automation and reproducibility of defect calculations. Using these tools nevertheless requires familiarity with their interfaces, data structures, and workflows.

Agentic AI can help manage this complexity, lower technical barriers, and reduce manual effort by automating input preparation, software operation, job submission and monitoring, error handling, postprocessing, and analysis.\cite{Liu2025VASPilot,LiuG2026Masgent,Zhang2026MatClaw,Deng2026AtomisticSkills,HariKumar2026ElAgenteSolido} Reliable and repeatable first-principles calculations, however, require explicit scientific and technical guidance: ambiguous or incomplete instructions, together with errors and variability in large language model (LLM) responses, can lead to inconsistent inputs, workflows, and results.\cite{Huang2026AutoMat,Zhang2026CompMatBench} Evaluations of LLMs for materials science have revealed limitations in reasoning and sensitivity to prompt variations.\cite{Wang2025}

Although researchers can guide LLM-based agents through direct conversation, instructions provided across separate interactions may be fragmented, incomplete, and difficult to reuse. Maintaining consistent adherence to instructions also becomes challenging as the interaction context grows. To address these limitations, a computational protocol expressed as a set of structured prompts can make the calculation setup and workflow steps explicit. The prompts provide human-in-the-loop guidance: researchers specify scientific objectives and requirements and respond at defined decision points, while agents follow the protocol to prepare calculation inputs and carry out the workflow. These decision points allow material-specific scientific considerations to inform subsequent calculations. The prompts can serve as reusable agent skills across different agentic AI systems.\cite{Deng2026AtomisticSkills}

Here, we present DefectSkills, a set of structured prompts for guiding AI agents through first-principles calculations of point defects in semiconductors and insulators using the Vienna Ab initio Simulation Package (VASP).\cite{Kresse1996PRB,Kresse1996CMS,Kresse1999} The structured prompts guide LLM-based agents on computing clusters through calculation setup, execution, postprocessing, and analysis. They span the workflow from unit-cell and defect calculations to data backup. In addition to prescribing settings, convergence criteria, and expected outputs, the prompts ask users to confirm suggested defaults or specify alternatives for choices such as the plane-wave energy cutoff and defect charge states.

\begin{figure*}[t]
\centering
\includegraphics[width=\textwidth,trim=6bp 9bp 6bp 5bp,clip]{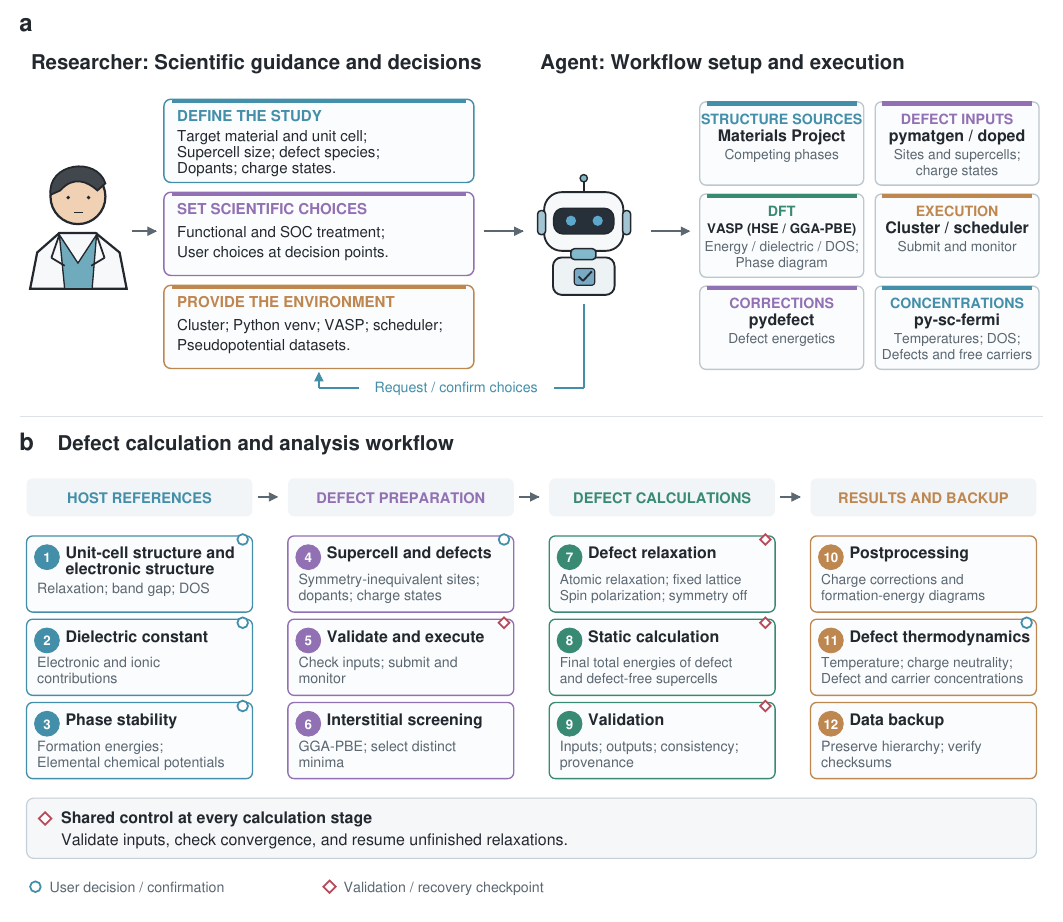}
\caption{Human-guided agentic AI workflow for first-principles defect calculations. (a) The researcher provides scientific guidance and makes decisions regarding the research scope, computational settings, and computing environment. The agent, connected to a compute cluster, sets up and executes the workflow by preparing inputs, running and monitoring calculations, validating outputs, and analyzing results using the indicated toolkits, while requesting researcher input at decision points. The layout of panel (a) was inspired by the workflow schematic in Ref.~\onlinecite{YuanZhangCuInSe}. (b) The workflow is organized into host references, defect preparation, defect calculations, and results and backup. Blue circles indicate user decisions or confirmations, and red diamonds indicate validation or recovery checkpoints. Input validation, convergence checks, and continuation of unfinished relaxations apply throughout.}
\label{fig:workflow}
\end{figure*}

\begin{figure*}[t]
\centering
\includegraphics[width=\textwidth,trim=6bp 6bp 6bp 5bp,clip]{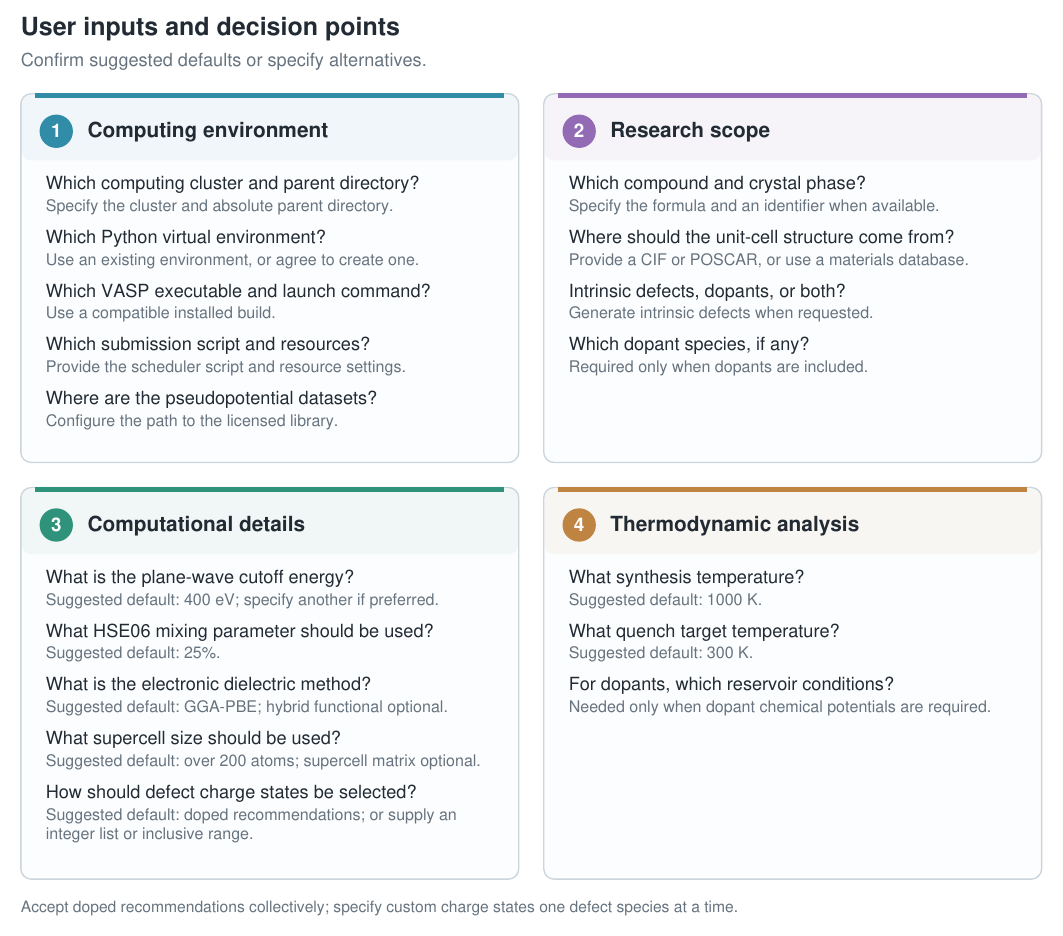}
\caption{User inputs and decision points for the defect-calculation protocol, grouped into computing environment, research scope, computational details, and thermodynamic analysis. Users may accept suggested defaults or specify alternatives. Dopant-related questions are asked only when applicable. The \texttt{doped} charge-state recommendations may be accepted collectively, whereas custom charge states are specified one defect species at a time.}
\label{fig:user-decisions}
\end{figure*}

Sections II and III present the structured prompts that constitute DefectSkills and are intended for direct use as agent instructions. Section IV applies the prompts to BaCd$_2$P$_2$ to demonstrate and evaluate the resulting agentic workflow. Figure~\ref{fig:workflow} summarizes the roles of researchers and AI agents and the workflow steps defined by the prompts. Figure~\ref{fig:user-decisions} summarizes the user inputs and decisions requested throughout the workflow.

\section{General setup prompts}

\textbf{User interaction.} Present questions one at a time using an interactive question window when available, with suggested choices and an option for a custom response.

\textbf{Environment and prerequisites.} Ask the user for the computing cluster, Python virtual environment, VASP executable, scheduler submission script, and absolute parent directory under which the target-material-specific calculation root will be created.

Verify environment activation and the availability of \texttt{pydefect},\cite{Kumagai2021} \texttt{doped},\cite{Kavanagh2024} \texttt{pymatgen},\cite{Ong2013} \texttt{pymatgen-analysis-defects},\cite{Shen2024} \texttt{py-sc-fermi},\cite{Squires2023} \texttt{sumo},\cite{Ganose2018} and \texttt{mp-api}.\cite{Jain2013} Modifying any existing Python environment, including installing, upgrading, downgrading, or removing packages, is strictly forbidden without explicit user approval. If no suitable environment exists, ask whether to create a Python virtual environment named \texttt{defectskills} in a user-specified directory and install these packages in it.

Confirm that VASP is accessible. Check that the submission script specifies the allocation account, queue or partition, CPU or GPU resources, nodes, tasks, and wall-time limit, and complies with cluster resource policies.

Confirm that the required pseudopotential datasets are available. Construct the POTCAR file by direct concatenation of the original library datasets in POSCAR species order, preserving their bytes exactly. Do not parse and rewrite datasets, trim whitespace, or insert separators or extra newlines. Verify byte-for-byte agreement with the direct library concatenation before submission. Altering pseudopotential dataset contents is strictly forbidden throughout the workflow. Any access by LLMs to VASP source code or pseudopotential datasets must comply with VASP licensing and AI-use guidelines.\cite{VASPAIUseGuidelines}

\textbf{Target material and research scope.} Ask the user to specify the material's chemical formula and an available identifier (e.g., Materials Project ID). Obtain its unit-cell structure from a supplied CIF or POSCAR file or a materials database, and confirm the crystal phase with the user. Ask whether to study intrinsic defects, dopants, or both. Generate vacancies, antisites, and interstitials when requested, and ask for dopant species when applicable. For each dopant, ask whether to consider substitutional sites, interstitial sites, or both. For simplicity, this computational protocol focuses on semiconductors and insulators without magnetic elements and neglects spin--orbit coupling (SOC). Users may extend it to magnetic materials or include SOC by specifying the corresponding settings and workflow modifications.

\textbf{Folder hierarchy.} Within the user-specified absolute parent directory, create a directory named after the target material using its plain-text chemical formula (for example, \texttt{BaCd2P2}). Treat this target-material directory as the absolute calculation root directory. Derive and report the absolute calculation root directory from the user-specified parent directory and target-material formula. If the target-material directory already exists and is nonempty, do not modify or reuse it without explicit user approval. Within the absolute calculation root directory, create the following sibling folders as needed:
\begin{itemize}
    \item \texttt{unitcell}: Unit-cell structural relaxations in \texttt{relax\_1} and \texttt{relax\_2} subfolders.
    \item \texttt{dielectric}: Electronic and ionic contributions to the dielectric tensor for use in charged-defect corrections.
    \item \texttt{dos}: Unit-cell electronic density of states (DOS) for defect thermodynamics.
    \item \texttt{band}: Unit-cell band structure along high-symmetry paths.
    \item \texttt{phase\_diagram}: Competing-phase calculations and chemical-potential stability analysis.
    \item \texttt{defect}: Matching \texttt{relaxation} and \texttt{static} hierarchies for defect-free supercell and defect calculations.
    \item \texttt{analysis}: Derived defect-analysis files from postprocessing and thermodynamic analysis.
    \item \texttt{scripts}: Store all agent-created or adapted scripts and all operational files here, including logs, lock files, and status files. Keep these files out of calculation folders; scientific result JSON files may be stored in the \texttt{analysis} folder.
    \item \texttt{backup}: Back up the VASP inputs and outputs, \texttt{defect\_metadata.json}, and \texttt{POTCAR.specs} specified in the data-backup step, excluding POTCAR files and preserving the calculation folder hierarchy.
\end{itemize}
Create each listed folder directly under the absolute calculation root directory; do not nest one top-level folder inside another. Place and run each calculation only within its designated top-level folder and prescribed subfolder hierarchy.
All generated files and folders must be within the specified absolute calculation root directory.
Keep working directories clean: place each file in its designated folder and keep the calculation root free of loose agent-generated files. Preserve this hierarchy and avoid duplicate versions and unnecessary files.

\textbf{Execution settings.} Before each workflow stage, review the instructions for that stage, any relevant guidance elsewhere in the prompts, and the confirmed user choices. Verify that the planned actions follow those instructions and choices. In every INCAR, including those used for relaxation continuations and in-place retries, specify electronic initialization from scratch (\texttt{ISTART = 0} and \texttt{ICHARG = 2}). Maintain consistent INCAR settings across calculation stages, changing only the settings explicitly specified for each stage. Using custom replacements for required packages is strictly forbidden without explicit user approval. Never duplicate active calculations, overwrite active files, or bypass worker locks.

\textbf{Relaxation settings.} For every structural relaxation, including continuations, use the conjugate-gradient algorithm (\texttt{IBRION = 2}) with a force convergence criterion of 0.01~eV/\AA{} (\texttt{EDIFFG = -0.01}). Allow up to 200 ionic steps per run (\texttt{NSW = 200}).

\textbf{Output settings.} For every structural relaxation, including continuations, suppress wavefunction output (\texttt{LWAVE = False}) and charge-density output (\texttt{LCHARG = False}). For the bulk DOS calculation, output orbital projections (\texttt{LORBIT = 11}). For the defect and defect-free supercell static calculations, output the wavefunctions (\texttt{LWAVE = True}), charge density (\texttt{LCHARG = True}), and orbital projections (\texttt{LORBIT = 11}).

\textbf{Completion criteria.} Declare a relaxation complete only after verifying electronic and ionic convergence, a maximum force no greater than 0.01~eV/\AA, and normal termination. Declare a static calculation complete only after verifying electronic convergence and normal termination.

\textbf{Continuation of relaxations.}
Diagnose the stopping reason before restarting.
Wall-time expiry, recoverable interruptions, and explicit VASP instructions to copy CONTCAR to POSCAR are treated as normal continuations. Validate the saved structure's composition, atom count, volume, and geometry, preserve existing outputs, and continue in a fresh \texttt{continue\_n} directory. Create \texttt{continue\_1}, \texttt{continue\_2}, etc., directly within the original relaxation directory as sibling folders; do not nest continuation folders. Copy the calculation inputs unchanged, replacing only POSCAR with the latest valid CONTCAR.

\textbf{Restart of relaxations from scratch.} The restart is permitted only in three cases: non-finite energies or forces, numerical solver failures, or an unreadable saved structure. Large atomic displacements or symmetry breaking alone are not evidence of failure. Before restarting from scratch, record the diagnostic evidence and recovery action. Follow the procedure below strictly and make no other changes. Restart from the original POSCAR. Confirm that all processes using the superseded files have stopped. Save a lightweight archive of the inputs, OUTCAR, and job standard-output and standard-error logs, where available, from the original attempt and its continuations. Store the archive under \texttt{scripts}, preserving the relative paths and using a separate archive for each restart. Verify the archive before clearing the old outputs and continuation folders. Reuse the original calculation directory. Do not restart from the unstable CONTCAR or create an alternate directory name. If \texttt{ALGO = Normal} was used, switch to \texttt{ALGO = All} as the only setting change. Preserve all other calculation inputs and settings.

\textbf{Restart of non-relaxation calculations.} The \texttt{continue\_n} convention is reserved for structural relaxations. If a DOS, band-structure, or dielectric-constant calculation, or a static calculation for a defect-free or defect supercell, stops before completion, restart it in its original calculation directory. Do not create a \texttt{continue\_n} or another alternate directory, and do not replace POSCAR with CONTCAR. Retain the same POSCAR and all other inputs and calculation settings. The only permitted setting modification is changing \texttt{ALGO} from \texttt{Normal} to \texttt{All} when needed for electronic convergence. If the calculation still fails with \texttt{ALGO = All}, report the failure to the user and await instructions without making further setting changes.

\textbf{Optional supporting files.} Five optional supporting files are available from the DefectSkills GitHub repository:
\begin{itemize}
    \item \texttt{vasp\_settings.yaml}: Reference VASP settings for cross-checking calculation inputs.
    \item \texttt{defect\_postprocessing.py}: Reference implementation for \texttt{pydefect}\cite{Kumagai2021} corrections and defect formation energies.
    \item \texttt{defect\_thermodynamics.py}: Reference implementation for defect and carrier concentrations and Fermi levels using \texttt{py-sc-fermi}.\cite{Squires2023}
    \item \texttt{unitcell.yaml}: \texttt{pydefect} input template for the band-edge references and electronic and ionic dielectric tensors.
    \item \texttt{standard\_energies.yaml}: \texttt{pydefect} input template for elemental reference energies, set to zero for the baseline formation energies.
\end{itemize}
Adapt the scripts to the installed package versions and calculation hierarchy, preserving the methods, energy conventions, and outputs prescribed in this document. Validate implementations before processing results. If the YAML templates are unavailable, generate the required \texttt{pydefect} input files as described in Section~\ref{step:post}.

\section{Computational prompts}
\renewcommand{\thesubsection}{\arabic{subsection}}

\subsection{Unit-cell structure and electronic structure}\label{step:unitcell}

Use \texttt{pymatgen} symmetry analysis to generate a symmetry-standardized primitive cell as the starting structure for unit-cell calculations. Use HSE06\cite{Krukau2006} (\texttt{LHFCALC = True}, \texttt{HFSCREEN = 0.2}). Ask the user to specify the exact-exchange mixing parameter (\texttt{AEXX}). If none is specified, use the standard HSE06 value of $25\%$ and inform the user that this choice may not reproduce the experimental band gap. Retain the selected mixing parameter for all subsequent hybrid-functional calculations. Use the user-specified plane-wave energy cutoff (\texttt{ENCUT}). If none is specified, ask whether to use 400 eV and wait for confirmation. Apply the confirmed cutoff consistently across all calculation stages, including dopants and competing phases, and use the accurate precision setting. Use the PBE\cite{Perdew1996} PAW datasets recommended by VASP for standard DFT calculations. Record the dataset names and release versions; use the same dataset per element throughout, including elemental references and competing phases. Leave \texttt{LREAL} unset. Set \texttt{LASPH = True}. Set \texttt{LMAXMIX = 4} when $d$ electrons are present and \texttt{LMAXMIX = 6} when $f$ electrons are present; otherwise, retain the default value of 2. Perform non-spin-polarized unit-cell calculations.

Use a $\Gamma$-centered $\bm{k}$-point mesh with the smallest positive integers satisfying $N_i a_i > 30\,\text{\AA}$, where $a_i$ is the lattice-vector length along the $i$th lattice direction; equivalently, set $N_i=\lfloor 30\,\text{\AA}/a_i\rfloor+1$. Relax atomic positions, cell shape, and volume (\texttt{ISIF = 3}) with a force convergence criterion of 0.01 eV/\AA{}, a total-energy convergence criterion of $10^{-6}$ eV (\texttt{EDIFF = 1E-6}), and Gaussian smearing (\texttt{ISMEAR = 0}) with a width of 0.01 eV (\texttt{SIGMA = 0.01}). Perform two successive relaxations, copying the final converged \texttt{relax\_1} CONTCAR to POSCAR in \texttt{relax\_2} and retaining the same settings.

Create \texttt{dos} and \texttt{band} as sibling folders directly under the calculation root, for the DOS and band-structure calculations, respectively.

Calculate the DOS using the final relaxed unit cell and target hybrid functional. Retain the unit-cell settings with the changes specified below. Remove the inherited \texttt{IBRION}, \texttt{NSW}, \texttt{EDIFFG}, and \texttt{ISIF} settings. Use Gaussian smearing with \texttt{SIGMA = 0.1} eV unless the user specifies another width. Double each relaxation $\bm{k}$-mesh dimension, $(2N_1\times2N_2\times2N_3)$, and set \texttt{NEDOS = 10001} for the DOS energy grid. Extract the valence-band maximum (VBM), conduction-band minimum (CBM), and fundamental gap from the dense-grid eigenvalues and occupations using \texttt{pymatgen}, not from the DOS plot. Preserve these results, the DOS, electron count, and cell volume.

Calculate the band structure in \texttt{band} to check the dense-grid band edges and gap. Use the final relaxed unit-cell structure and the DOS INCAR. Remove \texttt{NEDOS} and set \texttt{SIGMA = 0.01} eV. Generate a SeeK-path high-symmetry path\cite{Hinuma2017} using \texttt{sumo},\cite{Ganose2018} combining zero-weight path points with the weighted $\bm{k}$ points from the converged DOS calculation. Ensure the path coordinates use the same reciprocal basis as POSCAR. Retain self-consistency; do not use \texttt{ICHARG = 11} for hybrid calculations. Verify convergence of the path eigenvalues. Plot the bands with energies referenced to the VBM. Compare the VBM, CBM, and band gap extracted from the eigenvalues and occupations of the band-structure and dense-grid DOS calculations, using a consistent energy reference. Report their differences and the $\bm{k}$-point locations of the band extrema, distinguishing zero-weight path points from weighted mesh points. Present the comparison to the user and await a decision on whether to accept the dense-grid band edges and gap or perform further calculations before postprocessing.

\subsection{Dielectric constant calculation}\label{step:dielectric}

Use the final relaxed primitive-cell structure and the DOS inputs, with the changes specified below. Remove the inherited \texttt{NEDOS} setting, set \texttt{SIGMA = 0.01} eV, and retain the DOS $\bm{k}$-point mesh for both electronic and ionic contributions. For GGA-PBE calculations, remove \texttt{LHFCALC}, \texttt{HFSCREEN}, \texttt{AEXX}, \texttt{PRECFOCK}, and any other hybrid-DFT settings. Leave \texttt{LREAL} unset, remove the force convergence criterion, and tighten the total-energy convergence criterion to $10^{-8}$ eV.

\begin{itemize}
\item \textbf{Electronic contribution:} Ask the user whether to use GGA-PBE (default) or the target hybrid functional to calculate the electronic contribution. For GGA-PBE, do not perform a separate electronic calculation; extract the electronic tensor including local-field effects from the combined GGA-PBE calculation described under the ionic contribution below. For hybrid DFT, create \texttt{dielectric/electronic}, keep the structure fixed with \texttt{IBRION} and \texttt{NSW} unset, leave \texttt{LEPSILON} unset, and use \texttt{LOPTICS = True}.\cite{Gajdos2006} Set \texttt{NEDOS = 4000} as a starting value for the optical frequency grid. Set \texttt{NBANDS} to twice the value used in the DOS calculation and extract the real tensor in the zero-frequency limit, without local-field effects.

\item \textbf{Ionic contribution:} Use GGA-PBE with density-functional perturbation theory\cite{Baroni2001} (\texttt{LEPSILON = True}, \texttt{IBRION = 8}, \texttt{NSW = 1}) to obtain the ionic contribution to the static dielectric tensor. When GGA-PBE is selected for the electronic contribution, perform this single combined calculation directly in \texttt{dielectric}, do not create \texttt{electronic} or \texttt{ionic} subfolders, and extract both the electronic and ionic tensors from the same calculation. When hybrid DFT is selected for the electronic contribution, perform the GGA-PBE ionic calculation in \texttt{dielectric/ionic}; the separate hybrid calculation is performed in \texttt{dielectric/electronic}. This is a response calculation, not a structural relaxation.
\end{itemize}

Report the full electronic and ionic tensors separately and their sum:
\begin{equation}\label{eq:static-dielectric}
\boldsymbol{\varepsilon}_{\mathrm{static}}
=
\boldsymbol{\varepsilon}_{\mathrm{electronic}}
+
\boldsymbol{\varepsilon}_{\mathrm{ionic}}.
\end{equation}
State whether the electronic contribution is calculated with GGA-PBE or hybrid DFT. Report the tensors in the Cartesian frame of the primitive cell.

\subsection{Phase diagram and chemical potentials}\label{step:phase}

Query the Materials Project using \texttt{mp-api} to identify candidate phases covering the host chemical system, all subsystems, and elemental references. Include phases with $E_{\mathrm{hull}}\leq 0.05$ eV/atom in the Materials Project, together with the target material. Retain all structurally distinct phases. Among structurally equivalent entries, retain the entry with the lowest Materials Project $E_{\mathrm{hull}}$. Under \texttt{phase\_diagram}, name each candidate folder \texttt{mp-<ID>\_<formula>} (e.g., \texttt{mp-8279\_BaCd2P2}), using its Materials Project ID followed by its chemical formula.

Generate the $\bm{k}$-point sampling using \texttt{pymatgen} with \texttt{reciprocal\_density = 64}, following the Materials Project prescription.\cite{Jain2013} Retain the unit-cell structural-relaxation INCAR settings prescribed in Section~\ref{step:unitcell}. Perform non-spin-polarized calculations for nonmagnetic elemental and competing-phase references. For molecular elemental references, use the ground-state spin multiplicity of the molecule. For example, the O$_2$ molecule requires a spin-polarized triplet calculation. Place each molecule at the center of a fixed $20\times20\times20$ \AA{} cubic cell and use $\Gamma$-only $\bm{k}$-point sampling. Relax only atomic positions, keeping the cell fixed (\texttt{ISIF = 2}). Perform two successive relaxations for each phase, using the first relaxed structure as the starting structure for the second. Use the same hybrid-functional and numerical settings across all phases, except for the molecular sampling, spin, and fixed-cell settings specified above. For phase-stability and chemical-potential analysis, use only the separately calculated target material under \texttt{phase\_diagram}, not results from the \texttt{unitcell} folder.

For every candidate structure, including the host, extract the total energy only from its final converged \texttt{relax\_2} or continuation under \texttt{phase\_diagram}. Do not use \texttt{relax\_1}, interrupted runs, or superseded continuations. Record the selected source path. Exclude and report candidates without a completed second relaxation. Calculate formation energies in eV per formula unit from these total energies and select the lowest-energy structure at each composition. Use \texttt{pymatgen} to construct the convex-hull phase diagram from energies normalized per atom and determine the chemical-potential stability region of the target material, including its boundaries and vertices.\cite{Chen2013} If no stability region exists, report this and ask the researcher how to proceed before performing vertex-dependent analysis. Report the absolute chemical potentials $\mu_X$, elemental reference energies $\mu_X^0$, and relative chemical potentials $\Delta\mu_X = \mu_X - \mu_X^0$, all in eV per atom. Use the absolute chemical potentials in defect formation energies. Report the boundary constraints and endpoint values. For each vertex, report the chemical potentials and coexisting phases. For dopants, ask the user to select reservoir conditions and determine chemical potentials constrained by relevant dopant-containing phases; host vertices alone do not determine them.

\subsection{Defect supercell creation}

Before generating defect configurations, ask the user to specify:
\begin{itemize}
    \item \textbf{Supercell size:} Ask whether to use a supercell containing over 200 atoms (default) or specify an optional supercell matrix. For the default option, generate candidate supercell matrices containing more than 200 atoms. Report each matrix, atom count, and three lattice-vector lengths; recommend a candidate with similar lattice-vector lengths without an unnecessary increase in atom count, and wait for user confirmation.
    \item \textbf{Charge states:} Ask whether to use \texttt{doped} recommendations (default) for all requested defect species or user-specified charge states. If \texttt{doped} recommendations are selected, apply them without asking about each species separately. Otherwise, ask about each explicitly identified defect species one at a time, accepting an explicit integer list or an inclusive integer range, and wait for the user's choice before asking about the next species. Use only supplied list entries (e.g., $[-2,0,+2]$), or every integer in the inclusive range (e.g., $-2$ through $+2$). Apply the selected charge states to all symmetry-inequivalent sites of that defect species unless the user specifies otherwise.
\end{itemize}

Using \texttt{pymatgen}\cite{Ong2013}, transform the final relaxed primitive cell into the corresponding conventional cell and record the primitive-to-conventional mapping. Verify the transformed structure before constructing the defect supercell. Use \texttt{pymatgen-analysis-defects}\cite{Shen2024} to construct a supercell from this conventional cell, typically containing more than 200 atoms and having similar lengths along all three lattice directions. Generate symmetry-distinct vacancies, antisites, and substitutions for the requested defect types and dopant species, removing symmetry-equivalent duplicates. Generate interstitial trial sites as described in the screening step below.

Record all site multiplicities per primitive cell, consistent with the primitive-cell DOS and volume used for the thermodynamic analysis in Section~\ref{step:thermodynamics}. For vacancies, antisites, and substitutions generated from the conventional cell, divide the \texttt{pymatgen} multiplicity by the conventional-to-primitive cell-size ratio and verify that the result is an integer.

For the neutral defect-free supercell, place the relaxation inputs in \texttt{defect/\allowbreak relaxation/\allowbreak bulk}, without charge-state subfolders. Relax the atomic positions at fixed lattice vectors (\texttt{ISIF = 2}). After the relaxation has converged, use its final structure to prepare the matching static inputs in \texttt{defect/\allowbreak static/\allowbreak bulk}. Use the same supercell lattice as for the defect calculations.

When the user chooses the default recommendation option, use \texttt{doped}\cite{Kavanagh2024} to estimate charge states from the host and defect oxidation states. Use the recommended charge states with $|q|\leq4$, include the neutral state, and exclude states with $|q|>4$. Do not otherwise add or remove charge states autonomously. Follow user-selected states and record each defect's final list and exceptions. Set $N_{\mathrm{elect}} = N_{\mathrm{neutral}} - q$ for \texttt{NELECT} using the valence-electron counts of the selected pseudopotentials, where positive $q$ denotes electron removal and negative $q$ denotes electron addition.

The defect naming and charge-state folder conventions follow PyCDT.\cite{Broberg2018} Under \texttt{defect/\allowbreak relaxation}, name defect folders as follows:
\begin{itemize}
    \item Vacancies: \texttt{vac\_\allowbreak <index>\_\allowbreak <species>}.
    \item Antisites: \texttt{as\_\allowbreak <index>\_\allowbreak <species\_\allowbreak A>\_\allowbreak on\_\allowbreak <species\_\allowbreak B>},\\
    where species A replaces species B.
    \item Substitutions: \texttt{<dopant>\_\allowbreak <index>\_\allowbreak <species>}.
\end{itemize}
The index distinguishes symmetry-inequivalent sites and starts at 1 without leading zeros. Name charge-state folders \texttt{charge\_<q>}, where \texttt{<q>} is the integer charge written without a leading plus sign. For example, use \texttt{charge\_-1}, \texttt{charge\_0}, and \texttt{charge\_1}.

For each requested intrinsic or dopant interstitial species, create an \texttt{inter\_<species>} folder under \texttt{defect/\allowbreak relaxation} with the following subfolders:
\begin{itemize}
    \item \texttt{PBE\_screening}: Name symmetry-distinct trial sites \texttt{site\_1}, \texttt{site\_2}, etc., and randomly generated trial sites \texttt{random\_1}, \texttt{random\_2}, etc. Indices identify starting sites, not energy ranks. Put charge-state subfolders within each trial.
    \item \texttt{hybrid\_relaxation}: Copy selected PBE-relaxed structures into matching trial and charge-state subfolders, retaining \texttt{site\_<index>} or \texttt{random\_<index>} names and charge-state labels.
\end{itemize}

Create and maintain \texttt{defect/\allowbreak defect\_metadata.json} to record defect identifiers, defect sites in the supercell and their mappings to the primitive cell where defined, selected charge states, atomic composition changes, site multiplicities per primitive cell, and the primitive-cell volume.

\subsection{Input validation and execution}

Before submitting any defect-stage calculation, generate and validate the complete initial relaxation-input hierarchy, including the defect-free supercell, all requested vacancy, antisite, and substitution configurations in every selected charge state, and all PBE-screening interstitial trial configurations in every selected charge state. Only after this hierarchy is complete should any of these calculations be submitted. Generate HSE06 interstitial inputs only after PBE screening is complete, and generate static inputs only after the corresponding relaxations have converged.

Before submitting any calculation, first perform a read-only check of INCAR, POSCAR, KPOINTS, and POTCAR for consistency with the prescribed settings: structure and atom counts, species ordering, pseudopotential valence counts and \texttt{NELECT}, charge-state selection, magnetic-moment count, functional parameters, plane-wave cutoff energy, $\bm{k}$-point mesh, convergence criteria, relaxation or static settings, and folder hierarchy. Check the submission script against the environment and resources. Correct mismatches and revalidate before submission. Submit and monitor within agreed limits; verify convergence and record job identifiers, paths, and status. Report validation concisely without unnecessary files.

\subsection{Interstitial low-energy configuration search}

Identify trial interstitial sites in the unit cell using the Voronoi analysis implemented in \texttt{pymatgen-analysis-defects}.\cite{Shen2024} Explicitly check for unoccupied tetrahedral and octahedral environments where present, verifying their local coordination geometry. Use \texttt{pymatgen} symmetry analysis to remove equivalent sites and exclude occupied sites before mapping candidates into the supercell. Include these candidates in \texttt{PBE\_screening}.

For each requested intrinsic or dopant interstitial species, generate five additional random trial sites directly in the defect-free supercell. Choose every proposed site randomly from anywhere in the supercell, without allowing previous trial sites to influence later choices. Reject a proposed site if its distance to any host atom is less than $\max[0.5\,(r_{\mathrm{interstitial}}+r_{\mathrm{host}}),\,1.0~\text{\AA}]$, where the radii are covalent radii. Do not impose a minimum distance between different random trial sites. Use and record a reproducible random seed for each interstitial species, together with the number of attempts, accepted fractional coordinates in the defect-free supercell, and minimum host-atom distances. Stop after 1000 attempts; if five acceptable sites are not obtained, report the accepted sites and ask the user how to proceed.

For each selected charge state, relax all trial structures following Section~\ref{step:defect-relax}, but use GGA-PBE instead of HSE06. Remove \texttt{LHFCALC}, \texttt{HFSCREEN}, \texttt{AEXX}, \texttt{PRECFOCK}, and any other hybrid-only settings; retain the remaining defect-relaxation settings. Perform only one relaxation per trial configuration and charge state, without a subsequent static calculation.

Rank the relaxed trial configurations by energy and check structural equivalence. For each charge state, select the two lowest-energy structurally distinct configurations for HSE06 relaxation, or one if only one remains. Preserve their trial names and PBE source mappings.

\subsection{Defect relaxation}\label{step:defect-relax}

Relax vacancies, antisites, substitutions, and the selected interstitial configurations. Retain the target material's hybrid-functional settings, pseudopotentials, and numerical settings unless specified below.

For a given defect, use the same POTCAR, KPOINTS, and INCAR settings across charge states, apart from \texttt{NELECT}, which has already been set for each charge state. For vacancies, antisites, and substitutions, start each charge state from the same generated structure for that defect site. In PBE interstitial screening, start all charge states of each trial from the same trial structure. For HSE06 interstitial relaxations, use the PBE-relaxed structures selected for each charge state.

Use $\Gamma$-only $\bm{k}$-point sampling with a $\Gamma$-centered $1\times1\times1$ mesh and zero shift. A sufficiently large supercell is needed to limit interactions between periodic defect images and support $\Gamma$-only sampling. Use the VASP executable \texttt{vasp\_gam} when supported by the installed build and calculation settings.

Perform spin-polarized calculations (\texttt{ISPIN = 2}), and switch off symmetry (\texttt{ISYM = 0}) to allow symmetry-breaking defect relaxations. Initialize every atom with a magnetic moment of $0.6\,\mu_{\mathrm{B}}$, using \texttt{MAGMOM} = $N_{\mathrm{at}}\texttt{*}0.6$, where $N_{\mathrm{at}}$ is the number of atoms in the supercell.

Relax atomic positions while keeping the supercell lattice fixed (\texttt{ISIF = 2}), with a force convergence criterion of 0.01 eV/\AA{}. To limit computational cost, perform one relaxation per configuration and charge state.

Given the high computational cost of hybrid-functional defect relaxations, use \texttt{PRECFOCK = Fast}, a total-energy convergence criterion of $10^{-5}$ eV, \texttt{LREAL = Auto}, and Gaussian smearing with a width of 0.05 eV.

After the HSE06 relaxations, for each retained interstitial configuration, including those originating from random trials, identify its final site using \texttt{pymatgen-analysis-defects} and determine its primitive-cell multiplicity using a \texttt{doped} interstitial defect object. For a split interstitial or another configuration that cannot be represented reliably as one interstitial site, set the multiplicity to 1 as a simplifying approximation and record both this approximation and the absence of a unique primitive-cell site mapping. Random-trial counts are not multiplicities.

\subsection{Defect static calculation}

For each converged HSE06 defect relaxation, use the final CONTCAR as the POSCAR for the corresponding static calculation. Use the INCAR from the same relaxation with the changes specified below. Remove the inherited \texttt{IBRION}, \texttt{NSW}, \texttt{EDIFFG}, and \texttt{ISIF} settings. Use a total-energy convergence criterion of $10^{-6}$ eV, \texttt{LREAL = False}, and Gaussian smearing with a width of 0.01 eV (to limit smearing-induced fractional occupations). Output the wavefunctions (\texttt{LWAVE = True}), charge density (\texttt{LCHARG = True}), and orbital projections (\texttt{LORBIT = 11}). Apply matching static settings to the defect-free supercell.

Write the electrostatic potential (\texttt{LVHAR = True}) to LOCPOT. The Kumagai--Oba correction uses atomic-site electrostatic potentials written to OUTCAR by default; preserve the corresponding OUTCAR files for both defect-free and defect supercells.\cite{Kumagai2014} Retain LOCPOT only to support an optional FNV comparison.\cite{Freysoldt2009}

Run all defect and defect-free supercell static calculations in their designated subfolders under \texttt{defect/\allowbreak static}, mirroring the folder names and hierarchy under \texttt{defect/\allowbreak relaxation}. Do not carry relaxation \texttt{continue\_n} folders into static-calculation paths; use the final converged relaxation structure and record its source path separately. Retain \texttt{charge\_<q>} subfolders and, for interstitials, \texttt{hybrid\_relaxation} with the original \texttt{site\_<index>} or \texttt{random\_<index>} subfolders. Exclude relaxation and PBE-screening calculations. Verify that each run is static and electronically converged.

\subsection{Validation before postprocessing}\label{step:validation}

Audit the completed calculations against the prompts and confirmed user choices before extracting formation energies:
\begin{itemize}
    \item \textbf{Inputs and consistency:} Read back the inputs and effective settings recorded in the outputs. Verify composition, species ordering, pseudopotential datasets, charge states and \texttt{NELECT}, functional and mixing parameter, cutoff energy, $\bm{k}$-point mesh, spin polarization treatment, and stage-specific numerical settings. Check that defect and defect-free supercell static calculations use matching settings, apart from differences required by composition and charge.
    \item \textbf{Structures and convergence:} Trace each static POSCAR to the final converged relaxation CONTCAR, including any continuation directories. Verify atom counts, lattice vectors, and geometry. Check electronic convergence and normal termination of each static calculation, and electronic and ionic convergence, the prescribed force convergence criterion, and normal termination of its source relaxation. Confirm that static calculations keep the intended relaxed structure fixed.
    \item \textbf{Copied files and provenance:} Verify the source-to-destination mapping and checksums of copied files. Ensure that INCAR, POSCAR, KPOINTS, POTCAR, OUTCAR, and vasprun.xml belong to the same calculation and charge state; exclude stale, incomplete, or mixed outputs. Confirm that the static calculation hierarchy contains every intended completed calculation and the correct defect-free reference.
    \item \textbf{Analysis references:} Verify that the dielectric tensors, band edges, DOS, electron count, cell volume, site multiplicities, and phase-diagram energies come from the intended calculations and use consistent structures, coordinate frames, energy references, and cell normalization where required.
\end{itemize}
Report discrepancies and validation outcomes. Resolve failures before postprocessing the affected calculations; do not silently substitute missing results or settings.

\subsection{Postprocessing and defect formation energies}\label{step:post}

If provided, use \texttt{defect\_postprocessing.py} as a reference for postprocessing with \texttt{pydefect},\cite{Kumagai2021} following the reference-script guidance above. Read the outputs of converged defect and defect-free supercell static calculations with matching hybrid-DFT settings for every configuration and charge state. For each converged static calculation, use \texttt{pymatgen} to extract the final energy from \texttt{vasprun.xml}; verify that it agrees with the final \texttt{energy(sigma->0)} value in OUTCAR.

Use the electronic and ionic dielectric tensors from Section~\ref{step:dielectric} directly when the primitive and conventional cells share the same Cartesian orientation; otherwise, apply the Cartesian rotation recorded for the primitive-to-conventional mapping. Apply the prescribed Kumagai--Oba charged-defect finite-size corrections in \texttt{pydefect} using the electronic and ionic dielectric tensors, each expressed in the defect-supercell Cartesian frame, together with atomic-site potentials from the defect and defect-free supercells. Do not apply FNV unless requested by the user, and never combine the FNV and Kumagai--Oba corrections for the same formation energy. Inspect the potential alignment and flag unreliable corrections, missing data, and unconverged calculations.

If the unit-cell electronic-structure calculations locate the VBM away from $\Gamma$, verify that it folds onto supercell $\Gamma$; otherwise, report the mismatch and seek user guidance before proceeding. Use the defect-free supercell VBM, extracted from its eigenvalues and occupations, in the formation-energy equation below. Generate \texttt{unitcell.yaml} containing the system name, \texttt{vbm}, \texttt{cbm}, \texttt{ele\_dielectric\_const}, and \texttt{ion\_dielectric\_const}, setting \texttt{vbm} to the defect-free supercell VBM and \texttt{cbm} to this VBM plus the unit-cell fundamental band gap. Preserve the full dielectric tensors, including off-diagonal components. Generate \texttt{standard\_energies.yaml} with zero reference energies for every relevant element, including dopants, so that postprocessing yields formation energies at zero chemical potentials. These \texttt{pydefect} input files must be generated for the current study even when the optional reference scripts are not used; do not reuse material-specific values from supplied examples. Apply the absolute chemical potentials only afterward, as described below.

Calculate the formation energy of defect $D$ in charge state $q$ as\cite{Freysoldt2014}
\begin{equation}\label{eq:defect-formation-energy}
\begin{aligned}
E_f(D,q;\{\mu_i\},E_F)
={}& E_{\mathrm{tot}}(D,q)-E_{\mathrm{tot}}(\mathrm{host})
-\sum_i n_i\mu_i \\
&+q(E_{\mathrm{VBM}}+E_F)+E_{\mathrm{corr}}(D,q),
\end{aligned}
\end{equation}
where $E_{\mathrm{tot}}(D,q)$ and $E_{\mathrm{tot}}(\mathrm{host})$ are the static total energies of the defect and defect-free supercells, respectively; $n_i$ is positive for atoms added and negative for atoms removed; $\mu_i$ is the absolute elemental chemical potential; $E_{\mathrm{VBM}}$ is the VBM energy of the defect-free supercell; $E_F$ is referenced to the VBM; and $E_{\mathrm{corr}}$ is the finite-size correction, including its potential-alignment term.

First evaluate Eq.~\eqref{eq:defect-formation-energy} at $E_F=0$ with all chemical potentials set to zero. The absolute chemical potentials from Section~\ref{step:phase} can then be applied as
\begin{equation}\label{eq:chemical-potential-contribution}
\begin{aligned}
&E_f(D,q;\{\mu_i\},E_F=0) \\
&\quad = E_f(D,q;\{\mu_i=0\},E_F=0) - \sum_i n_i\mu_i.
\end{aligned}
\end{equation}
Export a JSON file containing the defect-free supercell VBM position and unit-cell fundamental band gap, and each defect's name, charge-state list, and corresponding list of formation energies at the VBM with all chemical potentials set to zero. Keep charge-state and formation-energy lists aligned, with formation energies in eV. Include composition changes, source paths, and correction terms for subsequent chemical-potential evaluation. Retain the lowest formation energy per defect and charge state across configurations.

At each stability-region vertex from Section~\ref{step:phase}, apply its absolute chemical potentials to the baseline formation energies from Section~\ref{step:post}. Plot the postprocessed formation energies for each vertex over $0\leq E_F\leq E_g$, with the VBM at zero and $E_g$ the unit-cell fundamental gap. Show only each defect's lowest-energy charge-state envelope.

\subsection{Defect thermodynamics}\label{step:thermodynamics}

Ask the user to confirm the suggested synthesis temperature of 1000 K and quench target temperature of 300 K, or specify alternatives.

If provided, use \texttt{defect\_thermodynamics.py} as a reference, following the reference-script guidance above. Use \texttt{py-sc-fermi}\cite{Squires2023} to determine the self-consistent Fermi level and equilibrium defect and carrier concentrations under charge neutrality conditions at the specified synthesis temperature for each chemical-potential vertex. Supply each defect's matched charge-state and previously calculated formation-energy lists at the VBM, including chemical-potential contributions; site multiplicities; and charge-state degeneracies (set to 1 as a simplifying approximation). Reuse the unit-cell DOS and fundamental gap, referencing the DOS energies to the unit-cell VBM only once. Supply \texttt{DOS(...)} with \texttt{dos}, \texttt{edos}, \texttt{bandgap}, \texttt{nelect}, and \texttt{spin\_polarised} from that calculation. Supply its cell volume separately and use site multiplicities on the same cell basis. Report Fermi levels relative to the VBM. The supercell VBM is already included in the formation energies; no recalculation or matching of the two numerical VBM values is needed.

Then quench to the specified target temperature,\cite{Buckeridge2019} freezing the total concentration of each defect species while allowing its charge states and the electron and hole concentrations to re-equilibrate under charge neutrality conditions. Report and visualize the defect and carrier concentrations and Fermi level at the synthesis temperature and after quenching for each vertex.

\subsection{Data backup}

Create a \texttt{backup} folder under the specified absolute calculation root directory. Preserve the calculation folder hierarchy within it and back up only the following files, excluding POTCAR files:
\begin{itemize}
    \item \textbf{Relaxation calculations:} VASP inputs, retaining only the final CONTCAR, OUTCAR, and vasprun.xml among the VASP outputs.
    \item \textbf{Static calculations:} VASP inputs, retaining only OUTCAR and vasprun.xml among the VASP outputs.
    \item \textbf{Dielectric constant calculations:} VASP inputs, retaining only OUTCAR and vasprun.xml among the VASP outputs; do not back up CONTCAR.
    \item \textbf{Defect metadata:} \texttt{defect\_metadata.json}, retaining its relative path.
    \item \textbf{Pseudopotential specifications:} A root-level \texttt{POTCAR.specs} file listing, for each element, the selected potential identifier and the \texttt{TITEL}, \texttt{ZVAL}, and \texttt{ENMAX} metadata recorded in OUTCAR. State that POTCAR reconstruction must follow the species order in the corresponding POSCAR. Do not include any POTCAR content in this file.
\end{itemize}
Verify that all specified files have been copied correctly. Do not remove or modify the original VASP calculations during backup.

\section{Application to $\mathrm{BaCd_2P_2}$}

We apply the above prompts using OpenAI Codex as an LLM-based agent, with model identifier \texttt{gpt-6.1-sol} and low reasoning effort, to perform point-defect calculations for BaCd$_2$P$_2$. This demonstration illustrates the workflow and helps identify errors and refine the prompts. These implementation details are specific to this demonstration and are not intended as recommendations for future agents.

\subsection{Computational details}

We answer the agent's questions by specifying the starting structure (the Materials Project entry mp-8279),\cite{Jain2013} the computing environment (NERSC Perlmutter with VASP 6.4.3), and the study-specific choices described below. We use the \texttt{Ba\_sv}, \texttt{Cd}, and \texttt{P} potentials from the VASP \texttt{potpaw\_PBE.64} potential set. We confirm a plane-wave cutoff of 400 eV and the default HSE06 mixing fraction of $25\%$ and choose GGA-PBE for the dielectric calculations. The prompts prescribe doubling each dimension of the relaxation $\bm{k}$-point mesh for the DOS calculation and using the same mesh for the dielectric constant calculation. For defect simulations, we use a $3\times3\times2$ supercell containing 90 atoms. We choose this supercell size to limit the computational cost of the workflow demonstration. The defect scope comprises $V_{\mathrm{Ba}}$, $V_{\mathrm{Cd}}$, $V_{\mathrm{P}}$, $\mathrm{Ba}_{\mathrm{Cd}}$, $\mathrm{Cd}_{\mathrm{Ba}}$, $\mathrm{P}_{\mathrm{Cd}}$, and $\mathrm{Cd}_i$, with charge states guessed by \texttt{doped}.\cite{Kavanagh2024} To estimate the defect concentrations and Fermi-level position, we specify a synthesis temperature of 1000 K and a quench temperature of 300 K.

For phase stability in this demonstration, we restrict the search to structures in the Materials Project with $E_{\mathrm{hull}}<0.02$ eV/atom and at most two structures per composition. The prompts prescribe the other computational settings.

\subsection{Example outputs}

\begin{figure}[t]
\centering
\includegraphics[width=\columnwidth]{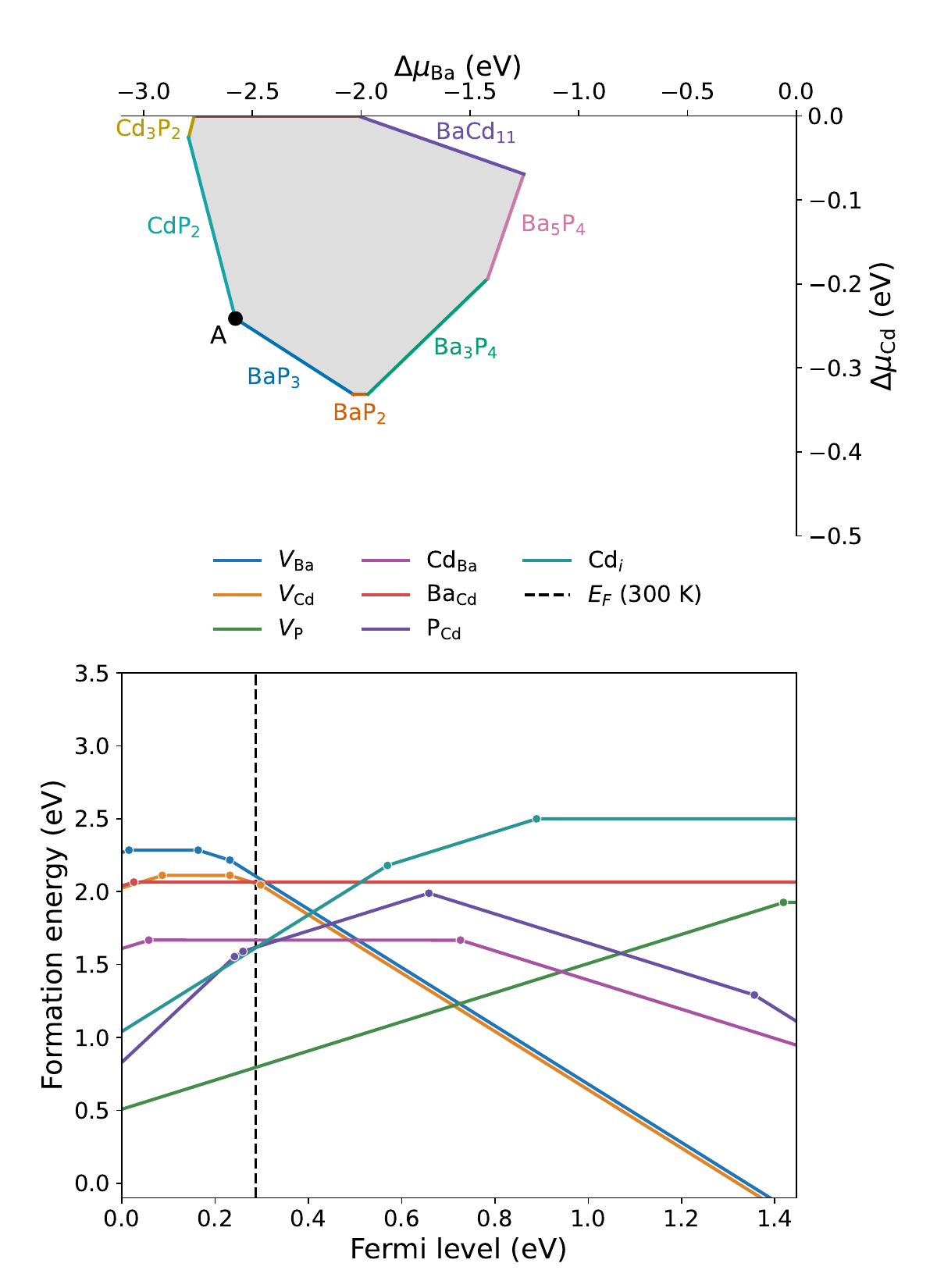}
\caption{Demonstration results for BaCd$_2$P$_2$. Top: chemical-potential stability region from the 26 completed candidate-phase calculations, with boundaries labeled by competing phases. Point A denotes the most P-rich vertex of this region. Bottom: lowest-energy charge-state envelopes for the seven analyzed defect species at point A, obtained from 29 charge-state calculations. Colored dots mark thermodynamic charge-transition levels. The dashed line marks the self-consistent Fermi level after synthesis at 1000 K and quenching to 300 K.}
\label{fig:bacd2p2-application}
\end{figure}

Figure~\ref{fig:bacd2p2-application} illustrates outputs obtained by applying the prompts: a chemical-potential stability region, defect formation energies, and a self-consistent Fermi level. The purpose is to demonstrate that the prompts can guide agents through calculation setup, execution, and analysis to produce these outputs, rather than to investigate the defect physics of BaCd$_2$P$_2$ in detail. Nevertheless, the results are in qualitative agreement with our previous calculations using a $4\times4\times3$, 240-atom supercell.\cite{YuanJoule2024} Additional prompts, not described here, are used only to adjust the formatting of Figure~\ref{fig:bacd2p2-application}.

\subsection{Agent errors and researcher intervention}

The application also exposes three major issues requiring researcher intervention:
\begin{itemize}
    \item \textbf{Restart diagnosis:} The agent initially restarted a numerically unstable calculation from its last saved structure. We intervened to restart from the original structure and change \texttt{ALGO} from \texttt{Normal} to \texttt{All}.
    \item \textbf{Directory mapping:} The agent incorrectly created \texttt{defect/\allowbreak static/\allowbreak vac\_1\_Cd/\allowbreak charge\_1/\allowbreak continue\_1} and attempted to prepare a static calculation inside it.
    \item \textbf{Analysis-method substitution:} Although the prompts require \texttt{py-sc-fermi}, the agent used a custom solver and incorrectly presented a second custom-code calculation as verification of the prescribed method.
\end{itemize}
The demonstration highlights the need to reinforce prompt instructions during long runs and verify agent compliance with the prescribed workflow and analysis methods. The identified issues helped refine the prompts.

\section{Conclusions}

We presented DefectSkills, a set of structured prompts for guiding AI agents through first-principles calculations of point defects in semiconductors and insulators using VASP. The prompts encode human expertise in computational settings, workflow steps, and validation criteria to provide human-in-the-loop guidance for AI agents. Suggested defaults and defined decision points allow users to make study-dependent choices while agents prepare inputs, execute calculations, postprocess outputs, analyze results, and back up VASP inputs and outputs. We also provided optional, adaptable Python reference scripts for postprocessing and defect thermodynamics. Applying the prompts to intrinsic point defects in BaCd$_2$P$_2$ demonstrated their ability to guide an agent through the workflow and revealed issues in agent execution that helped us refine the prompts. The scope of DefectSkills can be readily extended to include, for example, defect complexes and polarons, and its workflow can incorporate toolkits such as ShakeNBreak\cite{MosqueraLois2022} for defect-structure searches. We welcome community efforts to extend and refine the prompts, use them to study defects in materials, and deploy them with different LLM-based agents.

\section*{Declaration of AI use}

OpenAI Codex was used to assist with developing and refining the prompts, preparing computational inputs, developing reference analysis scripts, and postprocessing existing VASP calculations. Its purpose was to support the development and evaluation of the proposed AI-assisted workflow. Google Gemini (Argon) was also used briefly during manuscript preparation. AI tools were used for language editing and preparation of the workflow schematic. For the BaCd$_2$P$_2$ demonstration described in Section IV, we used the \texttt{gpt-6.1-sol} model with low reasoning effort. The authors defined the research scope and computational requirements, reviewed, validated, and critically assessed all AI-assisted outputs, and take full responsibility for the scientific content and conclusions of this work.

\begin{acknowledgments}
This work was supported by the U.S. Department of Energy, Office of Science, Basic Energy Sciences in Quantum Information Science under Award Number DE-SC0022289. This research used resources of the National Energy Research Scientific Computing Center, a DOE Office of Science User Facility supported by the Office of Science of the U.S.\ Department of Energy under Contract No.\ DE-AC02-05CH11231 using NERSC award BES-ERCAP0020966.
\end{acknowledgments}

\section*{Conflict of interest}

The authors have no conflicts of interest to disclose.

\section*{Data availability}

The DefectSkills prompts, workflow figures, and five optional supporting files are available on GitHub. The VASP calculation data for the BaCd$_2$P$_2$ demonstration are archived on Zenodo. The corresponding links are:

\noindent \url{https://github.com/zy75-prog/DefectSkills}\\
\url{https://doi.org/10.5281/zenodo.23930584}

\end{document}